# Synthesis of ferroelectric LaWN$_3$ - the first nitride perovskite

Kevin R. Talley[1,2], Craig L. Perkins[1], David R. Diercks[2], Geoff L. Brennecka[2,*], and Andriy Zakutayev[1,**]

[1]Materials Science Center, National Renewable Energy Laboratory, 15013 Denver West Parkway, Golden, Colorado, 80401, USA

[2]Department of Metallurgical and Materials Engineering, Colorado School of Mines, 1500 Illinois Street, Golden, Colorado, 80401, USA

[**]Andriy.Zakutayev@nrel.gov
[*]Geoff.Brennecka@mines.edu

**Abstract**: Next generation telecommunication technologies would benefit from strong piezoelectric and ferroelectric response in materials that are compatible with nitride radio-frequency electronic devices. Ferroelectric oxides with perovskite structure have been used in sensors and actuators for half a century, and halide perovskites transformed photovoltaics research in the past decade, but neither of them is compatible with nitride semiconductors. Nitride perovskites, despite numerous computational predictions, have not been experimentally demonstrated and their properties remain unknown. Here we report the experimental realization of the first nitride perovskite: lanthanum tungsten nitride (LaWN$_3$). Oxygen-free LaWN$_3$ thin films in a polar perovskite structure are confirmed by spectroscopy, scattering, and microscopy techniques. Scanning probe measurements confirm a large piezoelectric response and strongly suggest ferroelectric behavior, making it the first stable nitride ferroelectric compound. These results should lead to integration of LaWN$_3$ with nitride semiconductors for wireless telecommunication applications, while enabling synthesis of many other predicted nitride perovskites.

Nitride materials are revolutionizing the way humans access information and communicate with others.[1] For example, 4th generation (4G) wireless networks feature piezoelectric aluminum nitride (AlN) film bulk acoustics resonators (FBARs), and radio-frequency (RF) transistors based on semiconducting gallium nitride (GaN) are becoming an important part of the 5G telecommunication technology. The 5G infrastructure would also benefit from nitrides with ferroelectric responses (non-existent in AlN) and with much larger piezoelectric coefficients than AlN. Here we report on synthesis of lanthanum tungsten nitride (LaWN$_3$) with a perovskite crystal structure, large piezoelectric response, and ferroelectric switching behavior. LaWN$_3$ is the first ever synthesized oxygen-free nitrogen-rich perovskite material, and the first stable nitride ferroelectric compound. Synthesis of this new member of the broad family of natural and engineered perovskites (oxides, halides)[2], suggests that other computationally predicted nitride perovskites can also be realized. These nitride perovskites may also host a wide range of properties known from other perovskites and useful for numerous applications, such as nitrogen ion conductivity for sustainable electrochemical ammonia synthesis, defect-tolerant charge transport for solar energy conversion, giant anisotropic response for non-linear optics, or strongly correlated electrons for quantum information science.

Materials with the perovskite crystal structure (**Fig. 1a**) are the single most famous class of compounds that exhibit ferroelectric responses.(3) Ferroelectric oxide perovskites (e.g. Pb(Zr,Ti)O$_3$ (PZT), (Ba,Sr)TiO$_3$ (BST)) have been extensively used for ceramic capacitors(4), microelectromechanical actuators(5), and many other applications(6),(7) for the past century. In this decade, the number of reports on halide (X = Cl, Br, I) perovskites (e.g. CH$_3$NH$_3$PbI$_3$, CsPbI$_3$) has skyrocketed(8) because of their potential future application as inexpensive and efficient optoelectronic devices(**Fig. 1b**). Very recently, giant optical anisotropy and nonlinear optics applications attracted attention to chalcogenide (X = S, Se) perovskites (e.g. BaTiS$_3$, SrTiS$_3$).(9) In contrast to oxides, chalcogenides, and halides, there are no experimental reports of nitride perovskites in crystallographic databases or literature (**Fig. 1c**). One reported nitrogen-rich perovskite, (TaThN$_{3-x}$O$_x$) contained oxygen,(10) while other known nitrogen-poor and/or inter-metallic materials (e.g. Mg$_3$SbN, Mn$_3$CuN) have an anti-perovskite structure(11). This is very surprising, because pnictide (X = N, P) ABX$_3$ materials (including perovskites and others) are statistically more likely than the halide ABX$_3$ materials (**Fig. 1c**) due to a larger possible number of cation combinations that satisfy -9 versus the combinations for satisfying a -3 collective anion valence. *So how do we discover nitride perovskites and what properties would they have*?

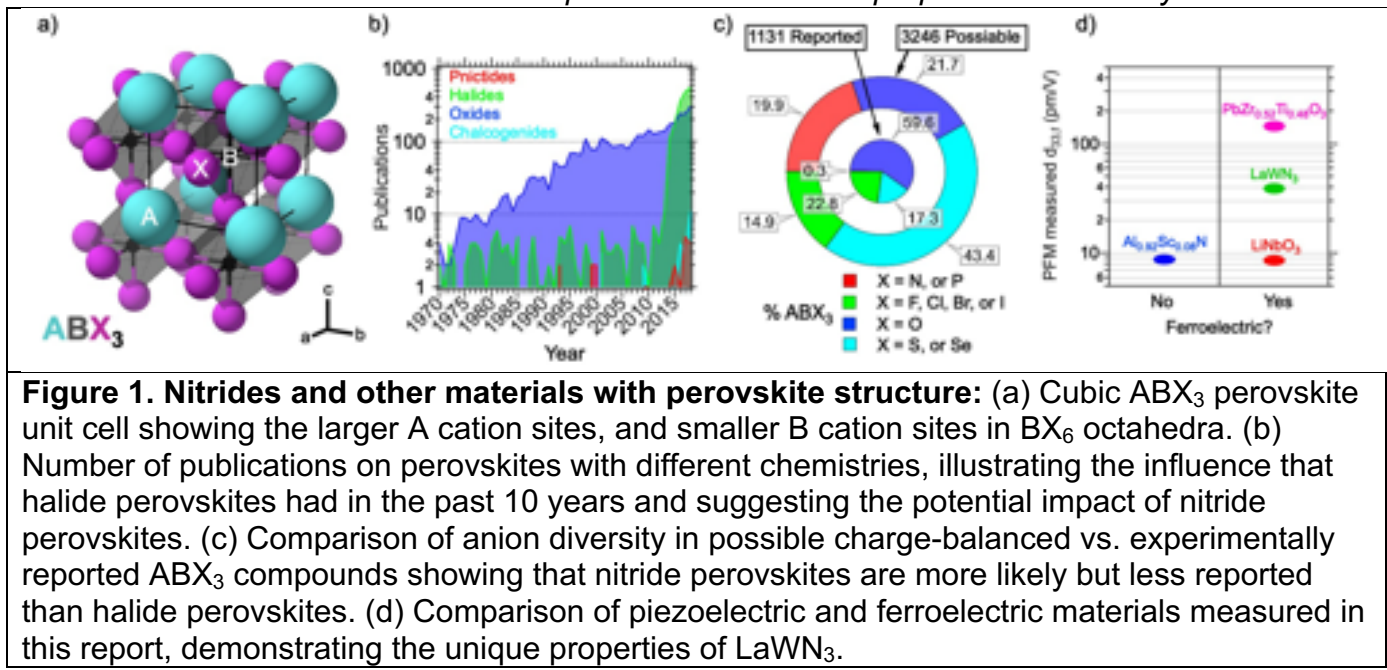

**Figure 1. Nitrides and other materials with perovskite structure:** (a) Cubic ABX$_3$ perovskite unit cell showing the larger A cation sites, and smaller B cation sites in BX$_6$ octahedra. (b) Number of publications on perovskites with different chemistries, illustrating the influence that halide perovskites had in the past 10 years and suggesting the potential impact of nitride perovskites. (c) Comparison of anion diversity in possible charge-balanced vs. experimentally reported ABX$_3$ compounds showing that nitride perovskites are more likely but less reported than halide perovskites. (d) Comparison of piezoelectric and ferroelectric materials measured in this report, demonstrating the unique properties of LaWN$_3$.

Computationally-driven experimental discovery is proving to be the most effective approach to predict and synthesize new materials.(12) In the field of nitrides, we have recently synthesized ~10 new ternary nitride materials(13) out of ~200 computational predictions(14). Among nitride perovskites, LnMN$_3$ (Ln=La, Ce, Eu, Yb, M=W, Re) materials were predicted by other groups to be stable(15),(16), with lanthanum tungsten nitride (LaWN$_3$) falling 350 meV/f.u. below thermodynamic stability. LaWN$_3$ is also predicted to have strong ferroelectric performance based on 61 µC/cm$^2$ spontaneous polarization magnitude with a small 110 meV barrier to polarization reversal.(17) However, synthesis of LaWN$_3$ and other oxygen-free nitrogen-rich perovskites by traditional bulk solid state chemistry methods remains extremely challenging,(18) often leading to oxynitrides (e.g., LaWO$_{0.6}$N$_{2.4}$).(19) If such nitride perovskites can be synthesized, a century of experience in perovskite property engineering(20), could be combined with decades of advances in nitride semiconductor integration(21) to have an enormous fundamental and applied impact.

Here, we report the synthesis of the first nitride perovskite, LaWN$_3$, with measured piezoelectric response much greater than any other known nitride, and ferroelectric response comparable to the well-known oxide perovskite compounds (**Fig. 1d**).

Crystalline LaWN$_3$ was synthesized by combinatorial physical vapor deposition on a heated substrate in ultrahigh vacuum with a nitrogen plasma source to ensure nitrogen incorporation and minimize oxygen contamination (see Methods in SI for more information). X-ray fluorescence spectroscopy (XRF) was used to measure the cation composition (La, W), and Auger electron spectroscopy (AES) depth profiling was used for the anion measurements (N, O) due to its highest relative sensitivity factor (RSF) to oxygen. As shown in **Fig. 2a-d**, no significant oxygen (below 4%) was detected throughout the thickness of films, even after 72 hours of atmospheric exposure, besides a thin (nm-scale) surface oxide layer. Scanning transmission electron microscopy (STEM) with energy dispersive x-ray (EDX) analysis (**Fig. 2e-h**) from the cross-section of an identical LaWN$_3$ film shows a polycrystalline microstructure (150-200 nm grain size) and demonstrate chemically homogeneity on the nanometer scale. The x-ray diffraction (XRD) patterns of the stoichiometric LaWN$_3$ thin films are consistent with the phase-pure perovskite reference pattern (**Fig. S.4c**), with W/WN and amorphous second phases at W- and La-rich compositions respectively (**Fig. S.4a**). Electrical and optical property measurements as a function of composition show $10^{-4}$-$10^4$ Ω cm resistivity (**Fig. S.5e**) and 1.0-2.5 eV optical absorption onset (**Fig. S.5f**) with increasing La content, with the upper bound being most accurate due to the optoelectronically inert character of the amorphous lanthanum oxide second phase.

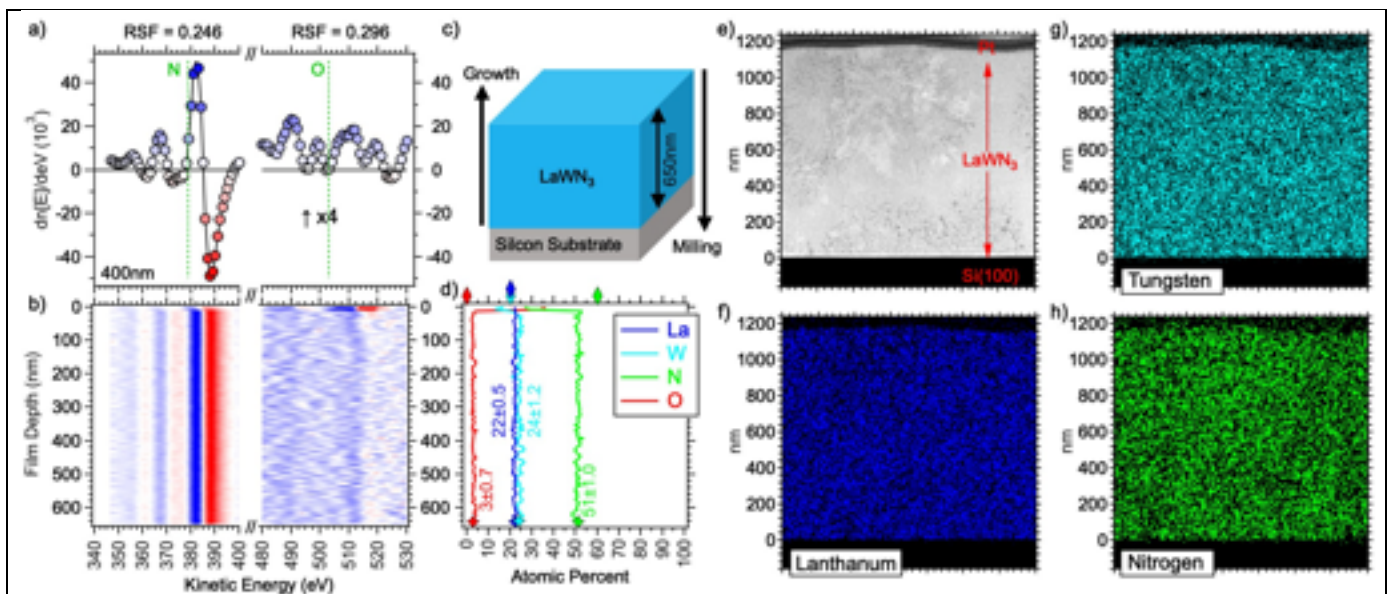

**Figure 2. Chemical composition of LaWN$_3$ thin films:** (a) Differentiated AES results and (b) depth-resolved color intensity map for O and N, showing negligible oxygen signal, even though AES is more sensitive to oxygen than nitrogen. (c) Depth profile and (d) resulting element concentration for all elements, with the average and ideal composition indicated by stars and diamonds at the bottom and top, respectively. (e-f) STEM and EDX images, showing a poly-crystalline microstructure and chemical homogeneity of LaWN$_3$ thin films.

To determine the crystal structure of LaWN$_3$, randomly-oriented polycrystalline thin films were synthesized by rapid thermal annealing (RTA) of atomically-dispersed La-W-N precursors deposited on glass substrates and protected from oxidation with an AlN capping layer (see SI for detailed methods). The capped amorphous sample libraries were also free of oxygen (**Fig. S.3**), and had a distinct color change close to the La/W=1 composition (**Fig. S.1**), from black on the W-rich side to translucent yellow on the La-rich side, which is consistent with a 2.5 eV band gap. Following RTA, a randomly-oriented polycrystalline microstructure is evident from uniform Debye-Scherrer rings (**Fig. 3a**). Rietveld refinement of the integrated XRD patterns (**Fig. 3b**) shows a majority LaWN$_3$ phase with a rhombohedral perovskite structure (R3c, space group 161) consistent with computational predictions(*17*) and a minority (5% by volume) metallic tungsten

(W) phase with a body-centered cubic (BCC) structure. The refinement, performed for the unit cell lattice vectors with all other variables held constant, of a tetragonal perovskite structure (I$\bar{4}$, space group 82)(*19*) resulted in low and statistically indistinguishable residuals (see **Fig. S.2**). STEM-based selected area electron diffraction (SAED) results (**Fig. 3c-f**) confirm the perovskite structure determined by XRD, and are also unable to resolve the I$\bar{4}$ (SG 82) vs. R3c (SG 161) structural distortion.

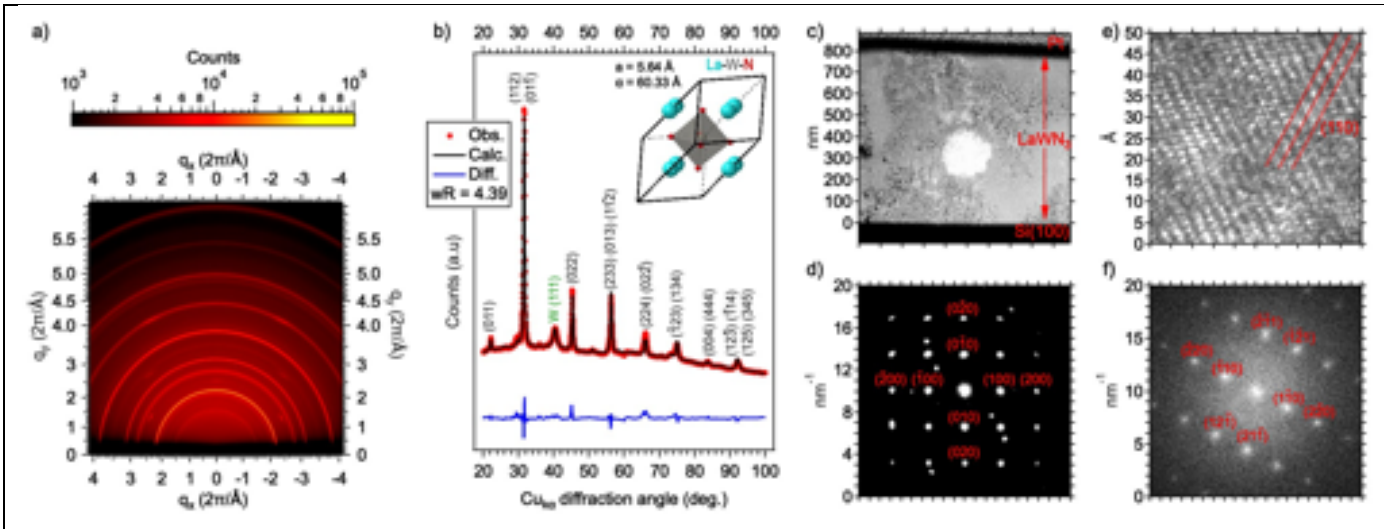

**Figure 3. Crystal structure of LaWN$_3$ thin films**. (a) Two-dimensional XRD pattern, indicating randomly oriented polycrystalline microstructure. (b) Rietveld refinements of XRD data to LaWN$_3$ with a rhombohedral unit cell (inset) of R3c symmetry (space group 161) and W minority phase (<5% by volume). (c) STEM-HAADF (high angle annular dark field) image of an as-deposited crystalline film highlighting a single grain (in white), and (d) SAED from this grain showing a pseudo-cubic perovskite [001] type pattern. (e) High resolution image of a single grain showing the pseudo-cubic (011) lattice spacing and (f) the associated fast Fourier transform of (e) indexed with a pseudo-cubic [113] type pattern.

In light of the computationally predicted piezoelectric properties and low barrier to ferroelectric switching of LaWN$_3$(*17*), piezoresponse force microscopy (PFM) measurements with a <25 nm tip radius were used to probe the electromechanical response and polar character of insulating (**Fig. S.6**) uncapped crystalline LaWN$_3$ thin films. The PFM results presented in **Fig. 4a-f** show an unambiguous piezoelectric response, and **Fig. 4g-h** suggest ferroelectric character of LaWN$_3$, notwithstanding PFM quantification issues discussed in literature(*22*). A map of effective piezoelectric strain coefficient $d_{33,f}$ (**Fig. 4d**) and statistical analysis of the measurement results (**Fig. 4f**) for LaWN$_3$ show a magnitude of response (~40 pm/V) that is larger than that of the Al$_{0.92}$Sc$_{0.8}$N (~10 pm/V) and LiNbO$_3$ (~10 pm/V) reference samples shown in **Fig S.10** and **Fig. S.7**. These PFM results clearly indicate a non-centrosymmetric unit cell, supporting the predicted polar R3c (SG 161) symmetry of LaWN$_3$, and ruling out centrosymmetric possibilities within 100 meV/f.u. from the ground state (See **Table S.3**).(*15*) Evidence of ferroelectric polarization reversal is presented in **Fig. 4g-h**, where the phase of the piezoelectric response of a single grain switches in the 0.25 - 0.50 MV/cm range. These values are similar to those measured for PbZr$_{0.52}$Ti$_{0.48}$O$_3$ (PZT) under identical conditions (**Fig. S.9**), meaning LaWN$_3$ is a comparable ferroelectric. While we are hesitant to claim quantitative values associated with piezoelectric coefficient and coercive field from these PFM measurements, the collective behavior presented in **Fig. 4** clearly demonstrates the piezoelectric nature of LaWN$_3$, and is highly suggestive of ferroelectric behavior.

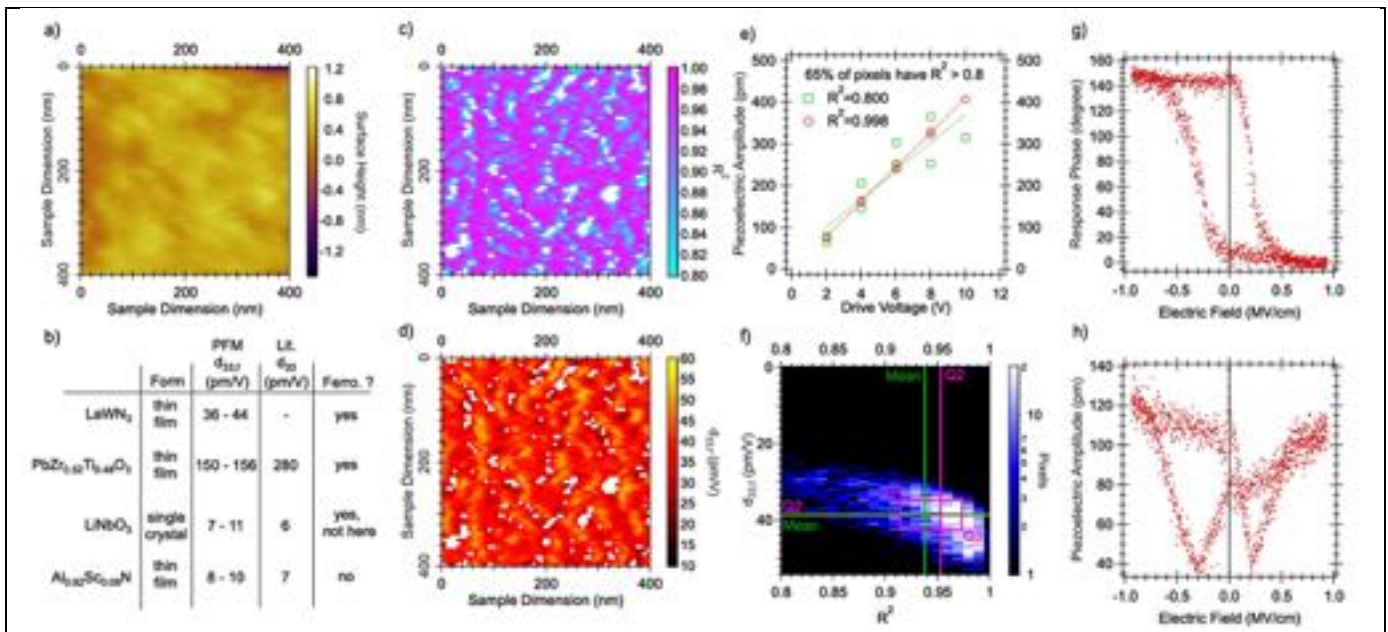

**Figure 4. Piezoelectric and ferroelectric properties of LaWN$_3$ thin films** (a) Atomically smooth surface of a single LaWN$_3$ grain and (b) comparison of other known piezoelectric and ferroelectric samples under the same analysis methods. (c) Linearity and (d) slope of each pixel with $R^2 > 0.8$ for piezoelectric amplitude vs. drive voltage fits. (e) The best and worst fits included in this analysis resulting in (f) a 3D histogram of all $d_{33,f}$ and $R^2$ values, indicating a piezoelectric response. (g) Hysteresis in piezoelectric phase and (h) piezoelectric strain response, consistent with ferroelectric character. See supporting information for analysis of LiNbO$_3$, PZT, and Al$_{0.92}$Sc$_{0.08}$)N reference samples and details of the PFM measurements (**Fig. S.7-10**).

It is very likely that piezoelectric and ferroelectric properties of LaWN$_3$ and other ABN$_3$ nitride perovskites can be improved using well-established design principles which have proven effective for oxide and halide perovskites. One promising avenue is increasing piezoelectric response in nitride perovskites using morphotropic phase boundaries, used in oxide-based piezoelectric perovskite research *(23)*. Well-known chemical strategies to achieve this goal include substituting A, B or X site species or imposing epitaxial strain to fine tune local and global distortions and resulting properties. One particularly intriguing option for future research is exploration of the boundary between the polar Rc3 LaWN$_3$ and non-polar I$\bar{4}$ LaWO$_x$N$_{3-x}$ with x=0.5 reported earlier.*(24)* However, such future piezoelectric and ferroelectric studies would benefit from deposition of textured LaWN$_3$ thin films without W or WN impurities, ideally on lattice-matched substrates, and macroscopic characterization of their piezoelectric and ferroelectric properties. Since both impurities are N-poor (W in 0 and +3 instead of +6 oxidation state), such results may be possible by further increasing the nitrogen plasma intensity during LaWN$_3$ synthesis (see SI for further discussion).

In the future, nitride perovskites could significantly extend the range of possible applications of existing commercial nitride semiconductor devices. GaN, AlN and related III-N alloys are well-established for electronics (e.g., radio-frequency transistor), photonics (e.g. light emitting diodes), and telecommunication (e.g. film bulk acoustic resonator). Thermodynamically stable ferroelectric nitride perovskite compounds may be easier to integrate with GaN than the recently reported metastable (Al,Sc)N alloy with high (>30%) Sc content*(25)*. Other computationally predicted nitride perovskites have useful properties such as record high anisotropy and large saturation magnetization in metallic TbReN$_3$, strong spin-orbit coupling and spin fluctuations in semiconducting EuReN$_3$(16), and large thermoelectric coefficient and topological insulator behavior in TaThN$_3$(26). Nitride perovskite are also likely to feature a variety of other emerging properties or hidden states, such as strongly correlated electrons(6) or quantum paraelectricity(7) known in oxide perovskites, but with even richer physics due to exceptionally strong metal-nitrogen bonds(18). Thus, integration of nitride perovskites with nitride semiconductors may lead to entirely new types of devices for emerging applications(21), as recently highlighted by examples of quantum computing and single-photon detectors in superconductor/semiconductor nitride heterostructures(27).

In summary, we report on the synthesis and characterization of the first oxygen-free nitride perovskite LaWN$_3$, which has been previously predicted by computation. LaWN$_3$ adopts the perovskite crystal according to Rietveld refinement of XRD data for randomly oriented polycrystalline LaWN$_3$ thin films crystalized by annealing of AlN-capped atomically-dispersed La-W-N precursors. LaWN$_3$ thin films deposited at elevated substrate temperatures are oxygen-free, compositionally-homogeneous and preferentially-oriented, which is suitable for property measurements. The results of PFM indicate a piezoelectric response greater than LiNbO$_3$ and strongly suggest ferroelectric behavior comparable to PZT. Integrating this LaWN$_3$ with modern III-N semiconductors would extend applications of nitride electronics in technologies such as emerging 5G wireless telecommunication infrastructure. This work also opens the door to synthesis of other predicted nitride perovskites with exceptional electromechanical, magnetic, optoelectronic, thermoelectric, topological, quantum, and other properties.

## Acknowledgements

This work was authored at the National Renewable Energy Laboratory, operated by Alliance for Sustainable Energy, LLC, for the U.S. Department of Energy (DOE) under Contract No. DE-AC36-08GO28308. Funding provided by Office of Science (SC), Office of Basic Energy Sciences (BES), as a part of the Early Career Award "Kinetic Synthesis of Metastable Nitrides" (synthesis and characterization); and by the U.S. National Science Foundation (NSF), Designing Materials to Revolutionize and Engineer our Future (DMREF) program, as a part of the "Computation of Undiscovered Piezoelectrics and Linked Experiments for Design" project DMREF-1534503 (piezoelectric and ferroelectric property measurements). Use of the Stanford Synchrotron Radiation Lightsource, SLAC National Accelerator Laboratory, is supported by the U.S. Department of Energy, Office of Science, Office of Basic Energy Sciences under Contract No. DE-AC02-76SF00515. The views expressed in the article do not necessarily represent the views of the DOE or the U.S. Government.

## Author contributions statement

K.R.T. synthesized the films, performed x-ray fluorescence and scattering measurements, piezoresponse force microscopy, and drafted the manuscript. C.P. performed Auger Electron Spectroscopy measurements and commented on the manuscript. D.D. performed transmission electron microscopy measurements and commented on the manuscript. G.B. provided intellectual guidance and physical resources, assisted in data analysis, and commented on the manuscript. A.Z. conceived the overall study, provided intellectual guidance and physical resources, assisted in data analysis, and edited the manuscript.

## Competing interests

The authors have no competing financial interest to disclose.

# Supporting Information

**Methods : Summary**

LaWN$_3$ thin films were deposited using reactive co-sputtering from elemental targets (La and W) on stationary Si and a-SiO$_2$ substrates. Depositions were performed in a vacuum chamber (10$^{-7}$ torr base pressure) with a nitrogen plasma source and a liquid nitrogen cooled cryogenic shroud surrounding the plasma zone for trapping residual water vapor to minimize concentration of oxygen. To further minimize oxygen impurities, the La targets were cleaned with hexane, loaded into the chamber in <1 minute, and pre-sputtered for 6 hours prior to use in deposition. Two deposition processes were developed and used for LaWN$_3$ synthesis with combinatorial La/W gradients across the (50 mm)$^2$ substrate. For producing polycrystalline materials for structural measurements, the films were deposited with no active heating, resulting in an amorphous structure, capped with AlN to protect from oxidation, and then thermally annealed for an hour at 900°C in flowing N$_2$ to crystallize. For producing materials more suited for property measurements, active heating of the substrate to 700°C during deposition led to crystalline materials with slight preferential orientation and did not require a capping layer. Additional details of these synthesis methods can be found in the supporting information.

Cation concentrations were measured using calibrated x-ray fluorescence (XRF), whereas anion and concentrations were measured using Auger Electron Spectroscopy (AES). This combination of methods allowed for the best resolution for heavy and light elements respectively. For structural characterization, wide-angle x-ray scattering (WAXS) was performed on beamline 11-3 of the Stanford synchrotron light source (SSRL) at the SLAC National Accelerator Laboratory and analyzed by COMBIgor and other data processing, fitting, and handling packages.(*1*),(*2*),(*3*) A lift-out specimen was prepared using a scanning electron microscope / focused ion beam instrument (SEM/FIB) and imaged using a transmission electron microscope (TEM). Energy dispersive x-ray spectroscopy (EDX) and high angle annular dark field (HAADF) imaging were performed in scanning TEM mode. Electromechanical properties of the material were investigated using conductive atomic force microscopy (c-AFM) and piezoresponse force microscopy (PFM), on the same instrument, using the same cantilever type. More information on characterization and analysis techniques are provided below.

**Methods : Film deposition**
Films were fabricated using a custom-built sputtering chamber with a base pressure of 10$^{-7}$ torr (measured prior to deposition). Elemental lanthanum (99.6%) and tungsten (99.9%) targets were held opposed from each other at 30° from substrate normal with a RF driven glow discharge throw distance of 7 cm. The lanthanum and tungsten targets were held at 0.74 and 0.37 W/cm$^2$ power densities respectively. Nitrogen (8 sccm) and argon (4 sccm) were kept at a total pressure of 4 mtorr during growth, where the nitrogen was introduced through an inductively coupled plasma source with 350 W of power, normal to the substrate from 15 cm away. The deposition environment was surrounded by a liquid nitrogen-filled shroud to minimize water as a source of oxygen contamination during growth. When the liquid-nitrogen shroud is turned on, the water signal in a residual gas analyzer is below the detection limit of 10$^{-9}$ torr. Films were deposited on both fused silica and p-type silicon substrates (50 mm)$^2$ with no active heating for 180 min following a 60 min pre-sputtering interval. No substrate rotation was used so the resulting films contained an intentional lateral gradient in lanthanum and tungsten concentration. Depending on the desired sample characteristics two temperature schemes were used. If a random polycrystalline film was desired, deposition with no active heating was followed by 60 min of AlN deposition (~100 nm) at a substrate temperature of 400°C to protect the amorphous film from oxidation upon removal from the growth chamber. Details on the AlN deposition conditions can be found elsewhere.(4) The film was then annealed at 900°C for 60 min with ramp rates of 1 °C/min under ultra-high purity $N_2$ in an ULVAC MILA-3000 to produce a fine-grained polycrystalline film. These conditions were used to investigate the structure and chemical compositions of $LaWN_3$. If a more textured film was desired, films were grown under identical conditions with active substrate heating to 700°C, with and without an AlN capping layer. These conditions were used to investigate the crystallinity, oxidation susceptibility, and electrical properties of crystalline $LaWN_3$.

**Methods : Composition measurements**

Compositions of the films were analyzed using a combination of Auger electron spectroscopy (AES) and x-ray fluorescence spectroscopy (XRF). The elemental composition of the cations was measured along the lanthanum/tungsten gradient by a Fischer XUV XRF instrument using a calibration based upon similar films of pure WN and pure LaN materials. Subsequent characterization was performed at film locations where the lanthanum and tungsten compositions were measured by XRF as equal. Select regions of the films were used to collect depth profiles on a PHI electronics AES 680 nanoprobe with a 5 kV/20 nA defocused electron beam, such that a circular area 50 microns in diameter was probed. Between measurement cycles, ion milling was performed using a 3 kV atomic argon beam. AES spectra collection resulted in more than 100 measurement cycles through the bulk of the $LaWN_3$ film. Under the conditions of the AES experiment, the oxygen limit-of-detection was estimated to be 3 percent. Details of the analysis method can be found elsewhere.(5)

**Methods : Structural measurements**

Wide angle x-ray scattering (WAXS) patterns were collected from the polycrystalline annealed film at the ideal 1:1 lanthanum:tungsten region. Measurements were conducted on beamline 11-3 of the Stanford synchrotron light source (SSRL) at the SLAC National Accelerator Laboratory. WAXS patterns were collected using a 12700 eV probe incident 87 degrees from substrate normal on a Rayonix MX225 CCD area detector placed 150 mm from the sample while the sample wobbled 0.5 mm in the plane of the substrate. The detector was calibrated using a $LaB_6$ standard and the Nika(3) package for Igor Pro, which was also used to dimensionally reduce the data to circularly average intensity vs. scattering vector data, which was converted from scattering vector to $Cu_{K\alpha}$ diffraction angle for structural refinement in the General Structure and Analysis System II software.(1) The variations in perovskite symmetry come from Ref. (6) and (7). WAXS patterns from the as-deposited crystalline films were also collected on beamline 11-3 at SSRL with the same collection parameters. Frames were taken at 1 mm steps across the cation gradient, and a $LaB_6$ calibration was applied. Data were managed and displayed using the COMBIgor package for Igor Pro.(2)

A site-specific lift-out transmission electron microscope (TEM) specimen was prepared using a Thermo Fisher Helios NanoLab 600i scanning electron microscope / focused ion beam instrument (SEM/FIB). Initial thinning was performed using a 30kV Ga ion accelerating voltage. The final surface cleaning step used 2kV. TEM analysis was performed on a Thermo Fisher Talos F200X instrument using a 200 keV accelerating voltage. Selected area diffraction patterns were collected using a selected area aperture having a projected size of approximately 200 nm. Scanning TEM operation was used for dark field imaging and compositional analysis via energy dispersive x-ray spectroscopy (EDX). For some dark field imaging, the high angle annular dark field detector was used with the camera length and specimen tilt adjusted to optimize the contrast of particular grains. For **Fig. 3f** the program ImageJ(8) was used to perform a fast-Fourier transform of the high-resolution TEM image in **Fig. 3e**.

**Methods : Property Measurements**

Conductive atomic force microscopy (C-AFM) was performed using an Asylum Research MFP-3D microscope with an ORCA dual gain holder and a platinum/iridium coated silicon cantilever. An area of the film was measured for electrical current, as low as nanoamps, under a forward and reverse bias condition as a means to test the insulating character of the uncapped film. Piezoresponse force microscopy (PFM) was performed using the same instrument as C-AFM with an additional Piezoresponse Force Module operating in a dual AC resonance tracking mode.(9) The platinum/iridium coated silicon cantilever was put in contact with the film surface near the stoichiometric $LaWN_3$ composition, and an area was measured for piezoelectric displacement amplitude, displacement phase, cantilever resonance frequency, and surface height under ramp-up and ramp-down drive amplitude bias conditions (0, 2, 4, 6, 8, 10 ,8 ,6, 4, and 2 volts). A linear fit of the piezoelectric displacement amplitude as a function of drive voltage was performed for each of the scan pixels, and the slope of each fit was taken as an effective piezoelectric coefficient with the fit $R^2$ taken as a measurement quality metric. An identical measurement and analysis of a periodically-poled $LiNbO_3$, PZT, and (Al,Sc)N reference samples was performed and is also presented in the supporting information as verification of the technique. Additional spectroscopy measurements were taken on select grains where a DC bias was applied in steps from -75 to 75 volts, through a full 20 cycles, and the piezoelectric response magnitude and phase were measured between the DC bias periods to detect changes in the piezoelectric response as potential domains reorient in response to the increasing DC field.

### Methods : Publication trends

The publication trends shown in **Fig. 1b** were retrieved from Clarivate Analytics Web of Science, 2019. A search was conducted with a filter for the word "perovskite" in the title, but not "anti-perovskite" or "inverse perovskite". For each of the three trends shown, one additional anion qualifier from the following sets was also included in the title. Pnictides had "nitride", "phosphide", or "pnictide" in the title, chalcogenides had "oxide", "sulfide", "selenide", or "chalcogenide" in the title, and halides had "fluoride", "chloride", "bromide", "iodide", or "halide" in the title. The publication record counts, per year, were downloaded and summed accordingly by chemistry. Then plotted from 1970 to 2018 for comparison. Year 2019 was excluded as the data are incomplete.

### Methods : Experimental reports of $ABX_3$ compounds

**Fig.1c** contains the distribution of experimentally demonstrated $ABX_3$ compounds from two databases, Open crystallography database(10) and PDF4+(11) All reported $ABX_3$ compounds were binned by the X anion: N, P, O, S, Se, F, Cl, Br, or I. Duplicate entries in both databases were removed and the total count used to calculate the percentages of experimental $ABX_3$ compounds shown in the figure. Using the above binning methods there is 1 $ABN_3$, 2 $ABP_3$, 675 $ABO_3$, 126 $ABS_3$, 70 $ABSe_3$, 86 $ABF_3$, 81 $ABCl_3$, 54 $ABBr_3$, and 37 $ABI_3$ compounds reported in these databases. None of the compounds were sorted based on reported structures, meaning it is likely that not all are perovskites. Because the combinatorics strategy outlined below does not consider the structure, structure is ignored here. This survey of reported $ABX_3$ compounds only considers mono-atomic site occupancy, meaning the poly-atomic ions commonly used in halide perovskite systems are not included.

### Methods : Combinatorics of $ABX_3$ compounds

**Fig. 1c** also contains the distribution determined by performing combinatorics with the ions found in the Shannon database(12). First, all ions were harvested from the database with corresponding oxidation states. A-sites ions came from groups 1, 2, 3, and the lanthanides. B site ions came from groups 4 - 15 excluding Au, Pt, and nonmetals. A sites were limited to oxidation states of 1-4 and B-sites limited to oxidation states of 1-6. All A-site and B-site cations considered are listed in this supporting information, **Table S.1** and **Table S.2** respectively. A script was written which then tried every possible A+B cation pair and binned the combinations by the total oxidation state of the pair. The total compounds, determined by combinatorics, were then just the total compounds with an oxidation state sum of 3, 6, or 9. Using this combinatorics strategy, it is determined there are 294 $ABN_3$, 294 $ABP_3$, 641 $ABO_3$, 641 $ABS_3$, 641 $ABSe_3$, 110 $ABF_3$, 110 $ABCl_3$, 110 $ABBr_3$, and 110 $ABI_3$ charge balanced compounds possible. This is a very simplified combinatorics strategy that ignores coordination and size of the cations, which are known factors of perovskite stability. However, by ignoring structure and focusing on charge balancing, a comparison can be made to the experimentally-reported compound diversity. This combinatoric $ABX_3$ analysis only considers mono-atomic site occupancy, meaning the poly-atomic ions commonly used in halide systems are not included.

**Table S.1.** Ions considered for A site perovskite occupancy in combinatorics calculations presented in the introduction section, mined from Shannon, R. D*., Acta Crystallogr. Sect. A*. **32**, 751–767 (2002).

| Ion | +1 | +2 | +3 | +4 |
|---|---|---|---|---|
| Ba |  | x |  |  |
| Be |  | x |  |  |
| Ca |  | x |  |  |
| Ce |  |  | x | x |
| Cs | x |  |  |  |
| Dy |  | x | x |  |
| Er |  |  | x |  |
| Eu |  | x | x |  |
| Gd |  |  | x |  |
| Ho |  |  | x |  |
| K | x |  |  |  |
| La |  |  | x |  |
| Li | x |  |  |  |
| Lu |  |  | x |  |
| Na | x |  |  |  |
| Mg |  | x |  |  |
| Nb |  |  | x | x |
| Nd |  | x | x |  |
| Pm |  |  | x |  |
| Pr |  |  | x | x |
| Rb | x |  |  |  |
| Sm |  | x | x |  |
| Sr |  | x |  |  |
| Tb |  |  | x | x |
| Tm |  | x | x |  |
| V |  | x | x | x |
| Yb |  | x |  |  |

.

**Table S.2.** Ions considered for B-site perovskite occupancy in combinatorics calculations presented in introduction section, mined from Shannon, R. D*., Acta Crystallogr. Sect. A*. **32**, 751–767 (2002).

| Ion | +1 | +2 | +3 | +4 | +5 | +6 |
|---|---|---|---|---|---|---|
| Ag | x | x | x | | | |
| Al | | | x | | | |
| B  | | | x | | | |
| Bi | | | x | | x | |
| Cd | | x | | | | |
| Co | | x | x | x | | |
| Cr | | x | x | x | x | x |
| Cu | x | x | x | | | |
| Fe | | x | x | x | | x |
| Ga | | | x | | | |
| Hf | | | | x | | |
| Hg | x | x | | | | |
| In | | | x | | | |
| Ir | | | x | x | x | |
| Mn | x | x | x | x | x | x |
| Mo | | | x | x | x | x |
| Nb | | | x | x | x | |
| Ni | | x | x | x | | |
| Os | | | | x | x | x |
| Pb | | x | | x | | |
| Pd | x | x | x | | | |
| Po | | | | x | | x |
| Re | | | | x | x | x |
| Rh | | | x | x | x | |
| Ru | | | x | x | x | x |
| Sb | | | x | | x | |
| Sn | | | | | x | |
| Ta | | | x | x | x | |
| Tc | | | | | x | |
| Ti | | x | x | x | | |
| Tl | x | | x | | | |
| V  | | x | x | x | x | |
| W  | | | | x | x | x |
| Zn | | x | | | | |
| Zr | | | | x | | |

**Methods : Possible perovskite distortions**

The polymorphs of LaWN$_3$ considered in this study are listed in **Table S.3**. There are some redundant structures found in the references. Space groups 221, 139, and 140 are the same by inspection of the La-lattice, the centering of the W-lattice, and the centering of the N-lattice. The I$\bar{4}$ polymorph is the reported structure of LaWO$_{0.6}$N$_{2.4}$ and is likely too large in volume as a result of the oxygen incorporation. The perovskite WN$_6$ octahedral network tiling ($a^0a^0c^-$) can also be supported by space group 140. Only R3c and R3m lack an inversion center to the unit cell symmetry and therefore have the ability to support a piezoelectric response. R3c is predicted to be 100 meV/f.u. lower in energy than R3m.

**Table S.3.** Table of LaWN$_3$ polymorphs. The energy values for each are in reference to the cubic Pm$\bar{3}$m space group.

| SG | # | Energy (meV/f.u.) | La-lattice parameter | W centered in La-lattice? | W centered in N$_6$? | Glazer tilting | Centrosymetric? | Ref. |
|---|---|---|---|---|---|---|---|---|
| I$\bar{4}$ | 82 | | 3.997,4.004 | yes | yes | $a^0a^0c^-$ | yes | (7) |
| Pm$\bar{3}$m | 221 | 212.22 | 3.964 | yes | yes | $a^0a^0a^0$ | yes | |
| I4/mmm | 139 | 211.71 | 3.962 | yes | yes | $a^0a^0a^0$ | yes | (6) |
| I4/mcm | 140 | 211.97 | 3.962 | yes | yes | $a^0a^0a^0$ | yes | |
| R$\bar{3}$c | 167 | 108.28 | 3.953 | no | yes | $a^-a^-a^-$ | yes | |
| Pnma | 62 | 106.29 | 3.961 | no | yes | $a^+b^-b^-$ | yes | |
| R3m | 160 | 104.35 | 3.975 | no | no | $a^0a^0a^0$ | no | |
| R3c | 161 | 0 | 3.967 | no | no | $a^-a^-a^-$ | no | |

**Results : Optical properties**

A transmission image of the as-deposited material on a fused silica substrate can be seen in the top of **Fig. S.1**. For comparison, the lanthanum cation fraction as measured by XRF is shown in the bottom of **Fig. S.1b**. The film transitions from opaque at approximately La/(La+W) = 0.47 to semi-transparent at about La/(La+W) = 0.53. **Fig. S.1b** shows the absorption coefficient as measured by a custom built UV-Visible light spectroscopy system with deuterium and tungsten halogen light sources. The measured spectra were used to calculate absorption coefficient as -Ln[T/(1-R)]/d, where d is the film thickness as measured by profilometry, and T and R are the fractions of transmitted and reflected light intensity respectively. In this region, the optical bandgap, determined by an absorption onset of $10^4$ cm$^{-1}$, is between 1.3-1.8 eV, although in the range of 0.55 < La/(La+W) < 0.6 the absorption onset begins to level off around 2.35 eV, indicating the band gap of LaWN$_3$ may be higher than shown here, in the amorphous state, when metallic bonds from a contaminant tungsten phase are not present.

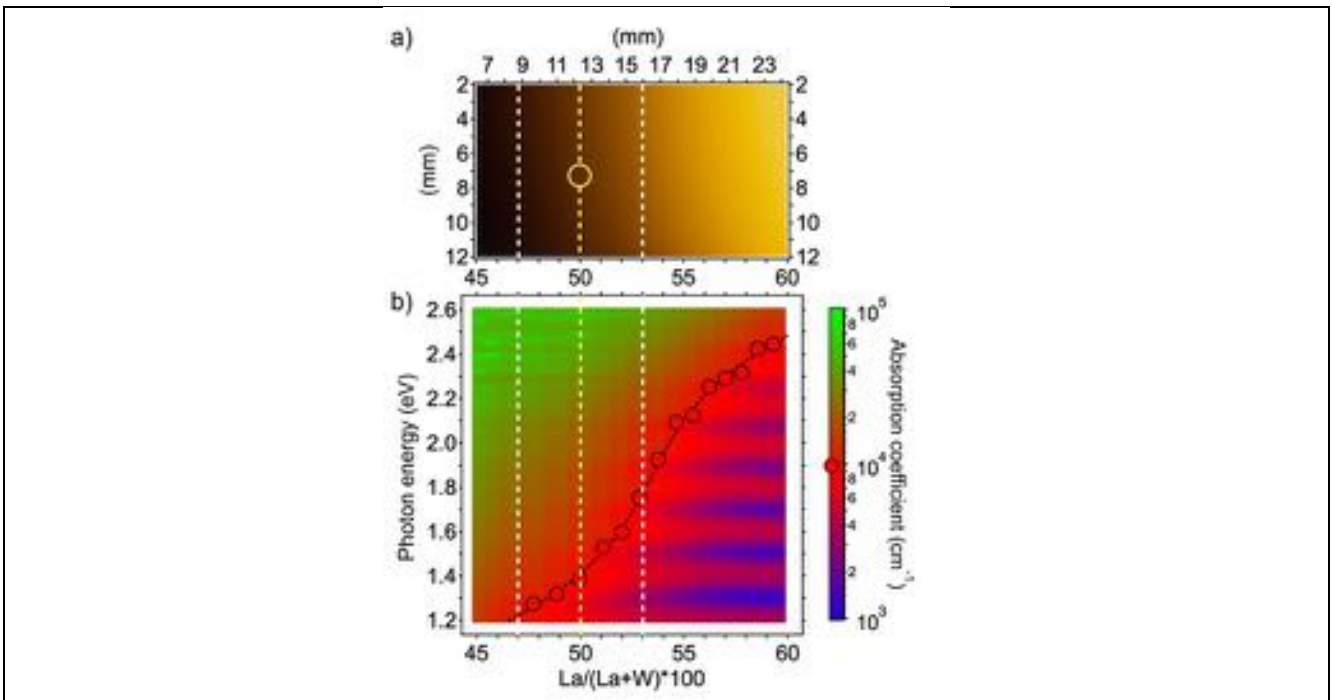

**Figure S.1.** (a) Transmission image of amorphous (La/W)N$_x$ film on a fused silica substrate and (b) corresponding optical absorption coefficient measured by UV-Visible light spectroscopy. A linear fit to La/(La+W) as a function of library dimension shows that LaWN$_x$ is at 12.6 mm (yellow dashed). The region between the two white dashed lines corresponds to a 3% change in lanthanum and tungsten concentration. The circles and trend line show an optical absorption onset of $10^4$ cm$^{-1}$, LaWN$_x$ has an optical band gap near 1.4 eV and a range of 0.5 eV over the 3% range.

# Results : Structural Refinement

To probe the structure of LaWN$_3$ films with a combination of WAXS and Rietveld refinement, films were deposited without active substrate heating, resulting in an amorphous structure, and then crystallized by thermal annealing in N$_2$ afterwards. The results of Rietveld refinement to R3c (space group 161) and I$\bar{4}$ (space group 82) structures are shown in **Fig. S2**.

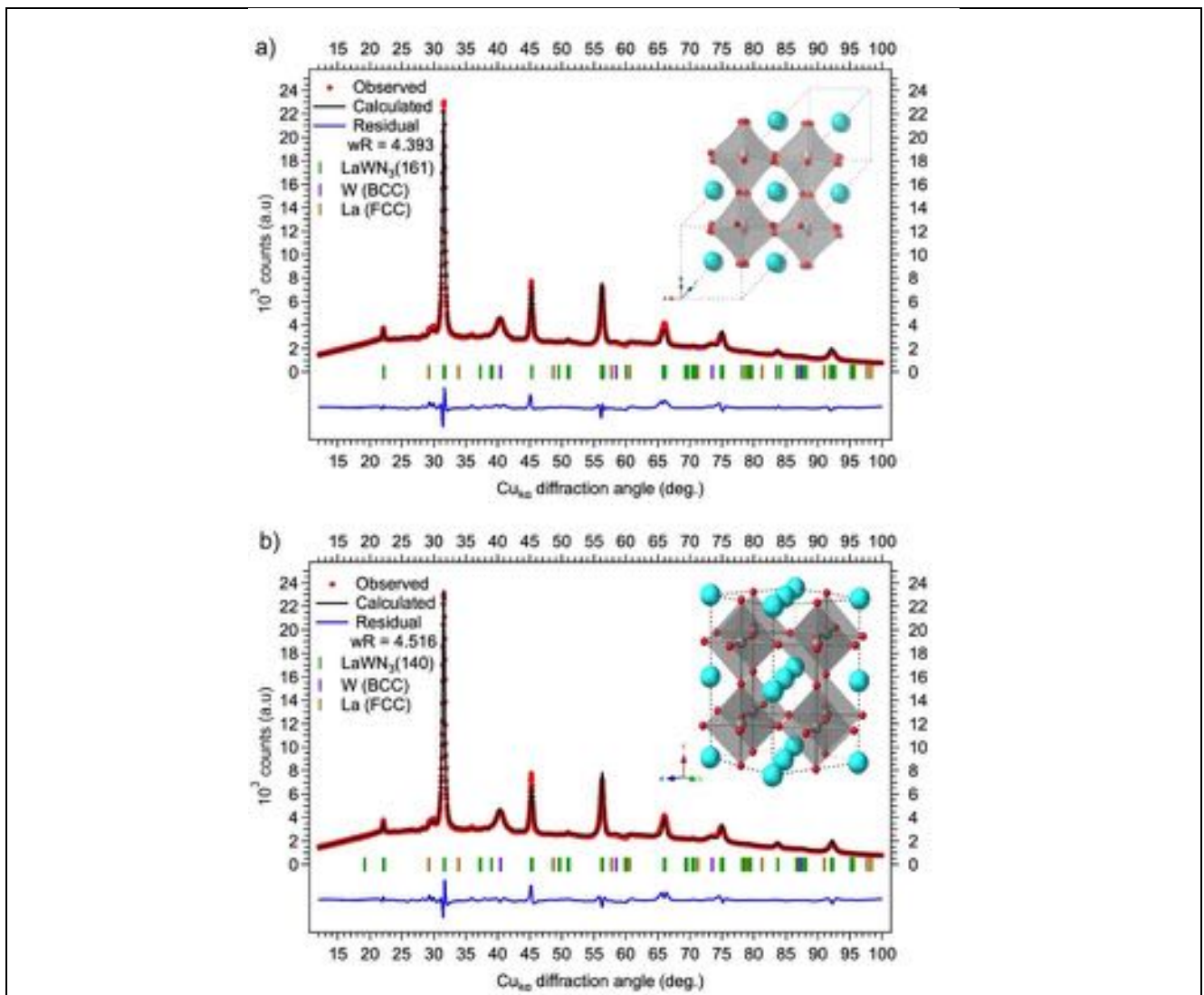

**Figure S.2.** Wide angle x-ray scattering (WAXS) pattern collected on the region shown in **Fig. S.1a** after thermal annealing to crystallize. Comparison between the (a) R3c (space group 161)(6) and (b) I$\bar{4}$ (space group 82)(7) choice of unit cell in GSAS refinement. The residuals (wR) are statistically equivalent, and no decision regarding which is more correct can be made by this method alone.

## Results : AES of As-deposited Amorphous Film

Crystalline LaWN$_3$ is oxidation resistant, whereas amorphous La-W-N films oxidize through the thickness and flake off the substrate within minutes of atmospheric exposure. To address this challenge, combinatorial La-W-N thin film sample libraries capped with AlN, with a La/(La+W) range of 0.45 to 0.60 across 15 mm, as determined XRF were produced for structural analysis. For the resulting amorphous material, with no detectable long range order according to laboratory scale x-ray diffraction, oxygen analysis of an amorphous films is presented in **Fig. S.3**. These AES results show an oxygen free material, similar to crystalline LaWN$_3$

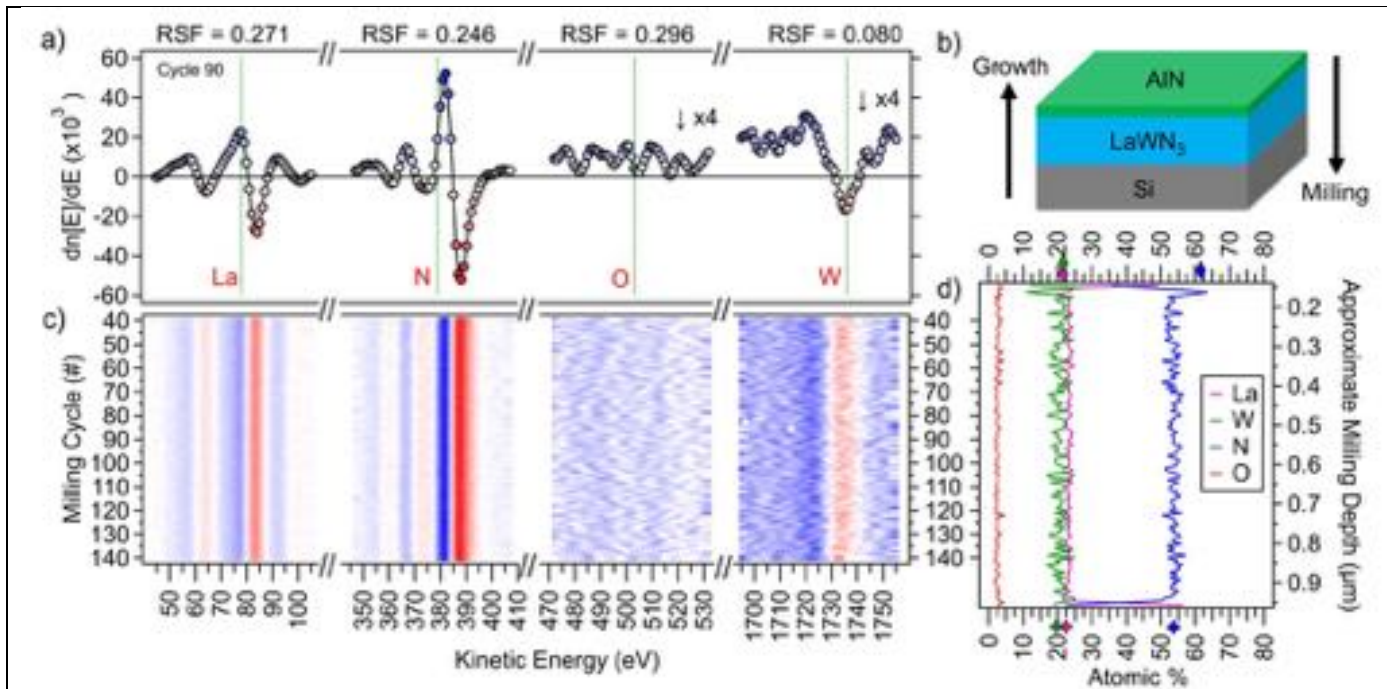

**Figure S.3.** Auger electron spectroscopy depth profile from the as-deposited amorphous film on a silicon substrate close to XRF stoichiometric cation composition. (a) Highlighted differentiated Auger electron spectra, with the machine relative sensitivity factors (RSF) at 5 keV probe energy and the most probable Auger electron energy for each element(13) as a dashed green line. The RSF is highest for oxygen, yet oxygen is below the detection limit within the LaWN$_3$ layer of the (b) film stack. Highlighted measurement-milling cycle in (a) is from the midpoint of the 100+ cycles through the LaWN$_3$ layer shown in (c). (d) Element concentration depth profile, with the first 150 nm of the film stack (30 milling cycles) not shown as they correspond to removal of the AlN capping layer. The average composition of cycles 40-140 is shown as stars on the bottom (22% La, 20% W, and 54% N), and the ideal stoichiometry is shown as diamonds on the top (20% La, 20% W, and 60% N).

## Results : Structural mapping

The perovskite phase persists across a wide range of La:W ratios, with metallic body centered cubic (BCC) tungsten or rocksalt (RS) WN(14) impurity phase at W-rich compositions and an amorphous impurity phase at La-rich compositions (**Fig. S.4a**). Peak distribution for in-plane wide-angle x-ray scattering (WAXS) in a narrow composition window close to stoichiometric La:W composition shows phase-pure perovskite material with a slight pseudo-cubic (110) preferential orientation (**Fig. S.4c**).

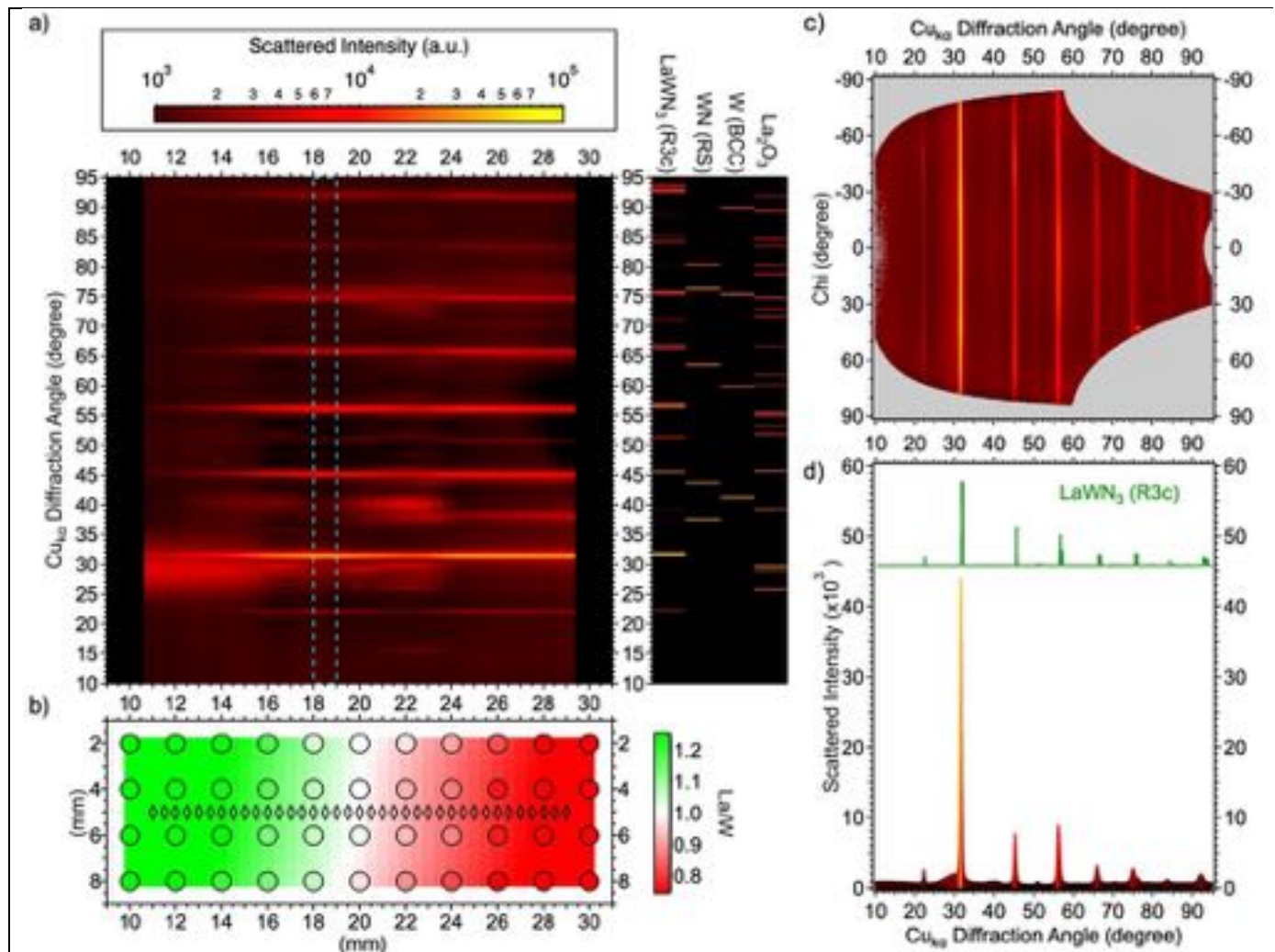

**Figure S.4.** (a) Integrated diffraction patterns across a compositionally-graded $LaWN_3$ thin film deposited crystalline with no AlN capping layer shows the signature of a perovskite structure at all La/W values where the (b) composition was measured with x-ray fluorescence (circles) and the scattering patterns (diamonds) were measured on BL 11-3 at SLAC SSRL. (c) Nika(3) calibrated detector for location closest to the ideal $LaWN_3$ composition, which is highlighted in (a). The c dimension is integrated to produce a (d) 2D profile where scattering vector magnitude was converted to a $Cu_{k\alpha}$ diffraction angle (2θ) for convenience.

## Results : Composite properties

Electrical and optical property characterization of these combinatorial sample libraries as a function of composition show resistivity values in the $10^{-4}$-$10^{4}$ Ω cm range (**Fig. S.5e**), and 1.0-2.5 eV optical absorption onset (**Fig. S.5f**). Both of these macroscopic optoelectronic properties may be approximately accurate for La-rich compositions but are likely affected by BCC-W or RS-WN impurities at W-rich compositions.

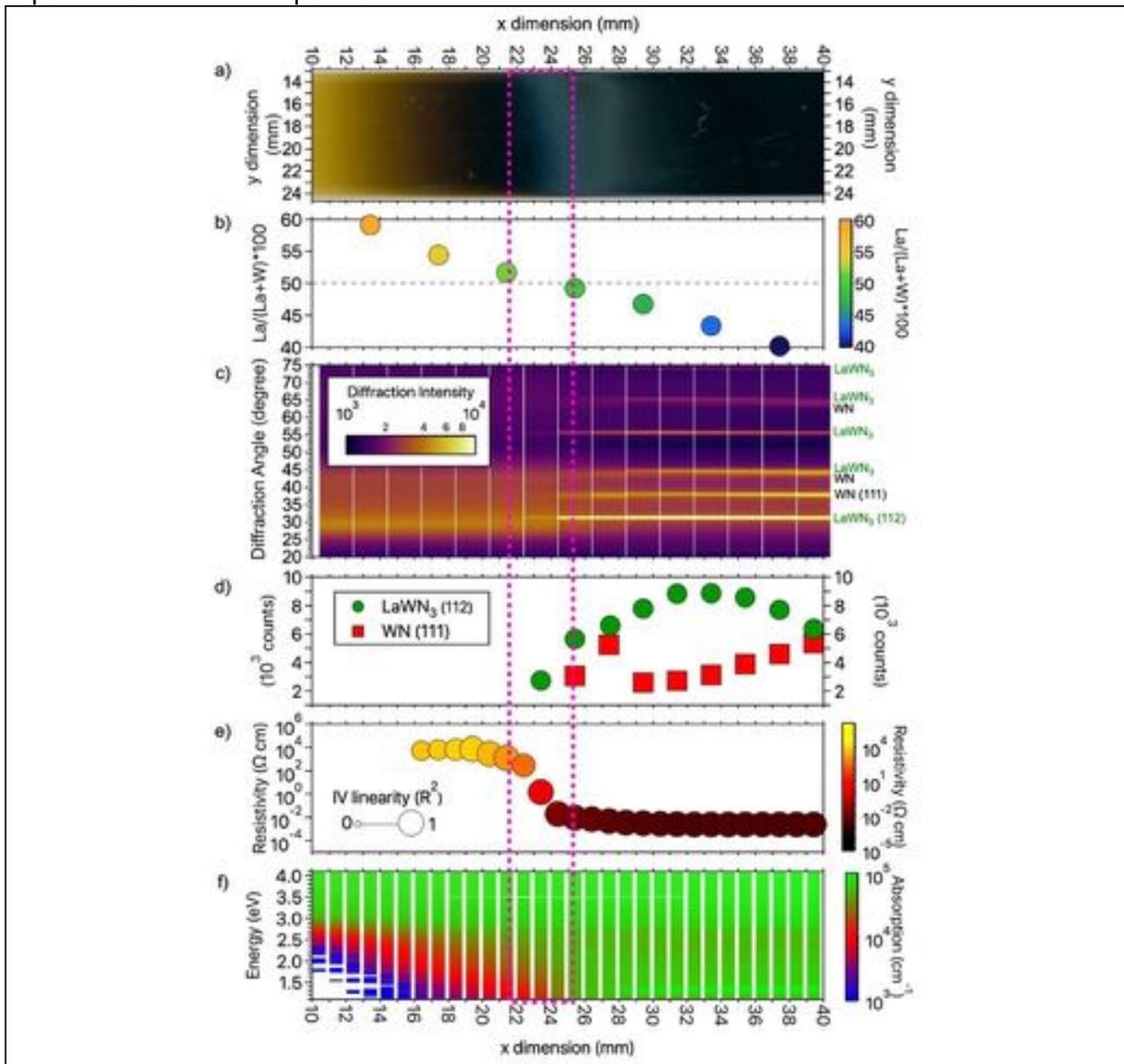

**Figure S.5.** Characterization and property measurements across a single compositionally graded (La/W)N$_3$ library deposited onto a fused silica substrate under active heating, resulting in crystalline material of mixed phases. The dashed lines indicate the region near the desired composition. (a) Transmission visible light image, (b) cation composition as measured by x-ray fluorescence, (c) laboratory scale x-ray diffraction patterns showing R3c-LaWN$_3$(*6*) and Rocksalt-WN(*14*) indexing, (d) integrated peak counts for the primary WN and LaWN$_3$ peaks, (e) four point probe electrical measurements where probes were in a linear configuration perpendicular to the composition gradient, and (f) UV-Visible light spectroscopy measured absorption coefficient.

**Results : Conductive atomic force microscopy**

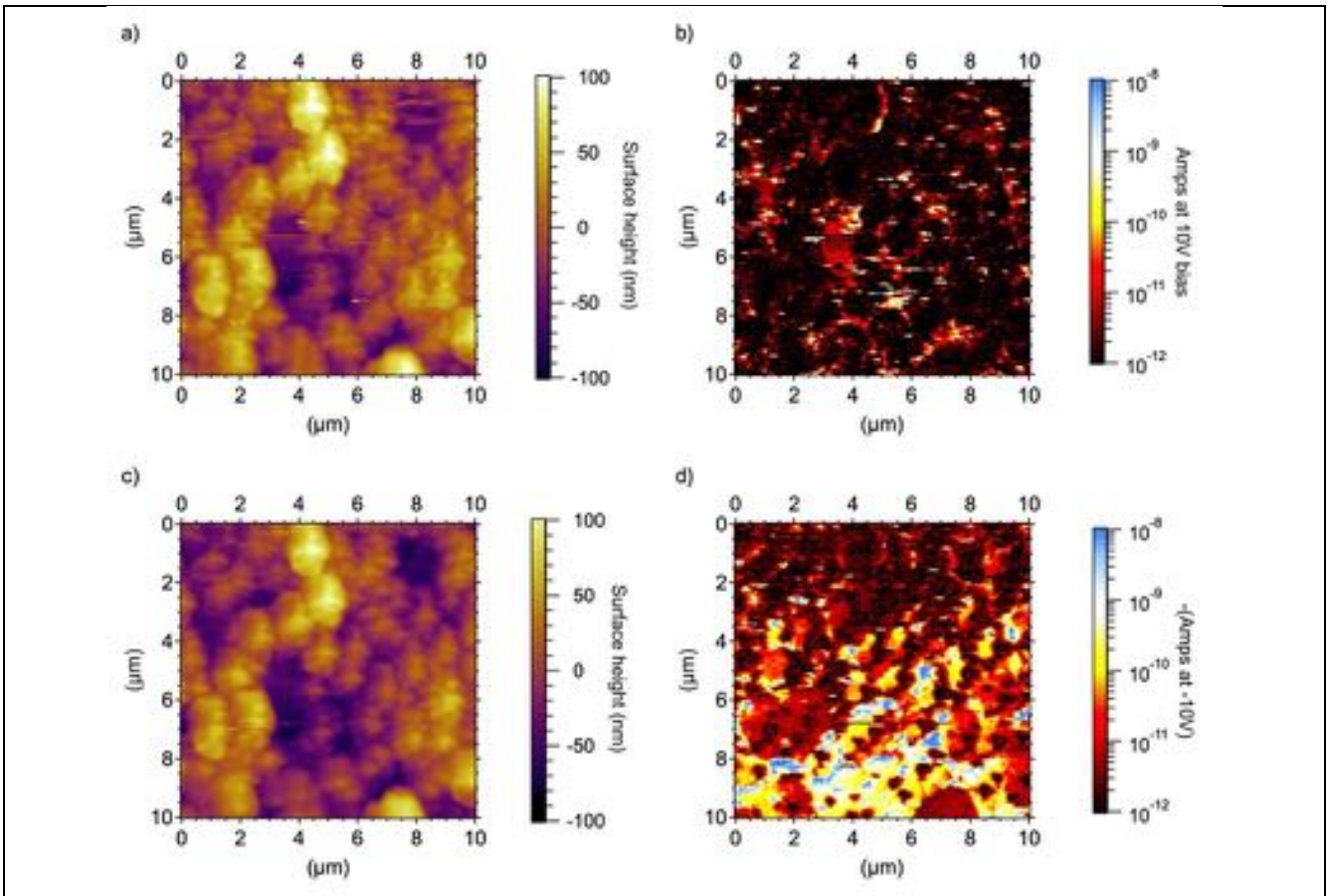

**Figure S.6.** Conductive atomic force microscopy results from an as-deposited crystalline LaWN$_3$ thin film. The surface structure observed in the height signal (a & c) is similar to that seen in **Fig. 5g**. The surface is essentially identical under a (a) forward and (c) reverse bias. The current conducted through the Pt coated silicon tip was primarily below 1 nA under a (b) 10V forward bias and below 10 nA (d) a 10V reverse bias. A 10V bias results in an approximate 2*10$^5$V/cm electric field strength. This analysis indicates that the crystalline film is essentially insulating under these conditions (-10V to 10V).

**Results : Piezoresponse force microscopy of periodically poled LiNbO₃ reference sample**

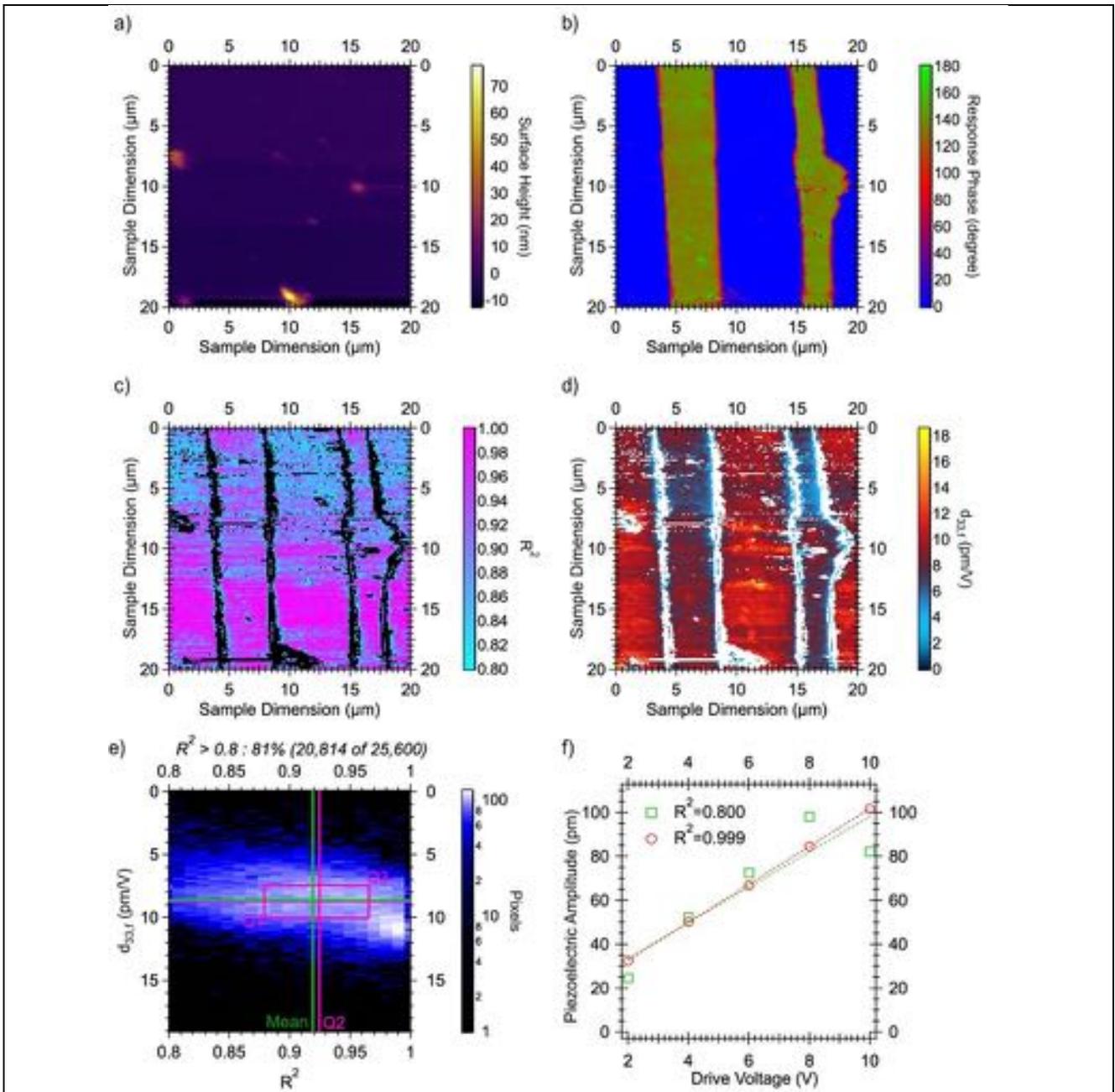

**Figure S.7.** Piezoresponse force microscopy results from periodically poled LiNbO$_3$ (PPLN) single crystal sample (https://afmprobes.asylumresearch.com/accessories/ar-ppln.html). (a) The surface is observed to be smooth from the height signal other than a few dust particles. Piezoelectric displacement and (b) response phase as a function of drive voltage was collected in a dual AC resonance tracking mode(9). Each pixel was fit with a line function to extract the (c) response linearity ($R^2$) and (d) effective piezoelectric strain coefficient ($d_{33,f}$) for each of the >25k pixels where only those with $R^2>0.8$ (81%) were analyzed further. The (d) slope and (e) linearity of each fit is shown for pixels where $R^2>0.8$. (e) A 3D histogram of all $d_{33,f}$ and $R_2$ values shows the statistical mean (green) and quartiles (magenta) of the pixel distribution. (f) Example pixel fits for the highest and lowest acceptable $R^2$ values. The results indicate PPLN has a clear piezoelectric response and reinforce the strength of this analysis method.

**Results: Ferroelectric testing DC bias**

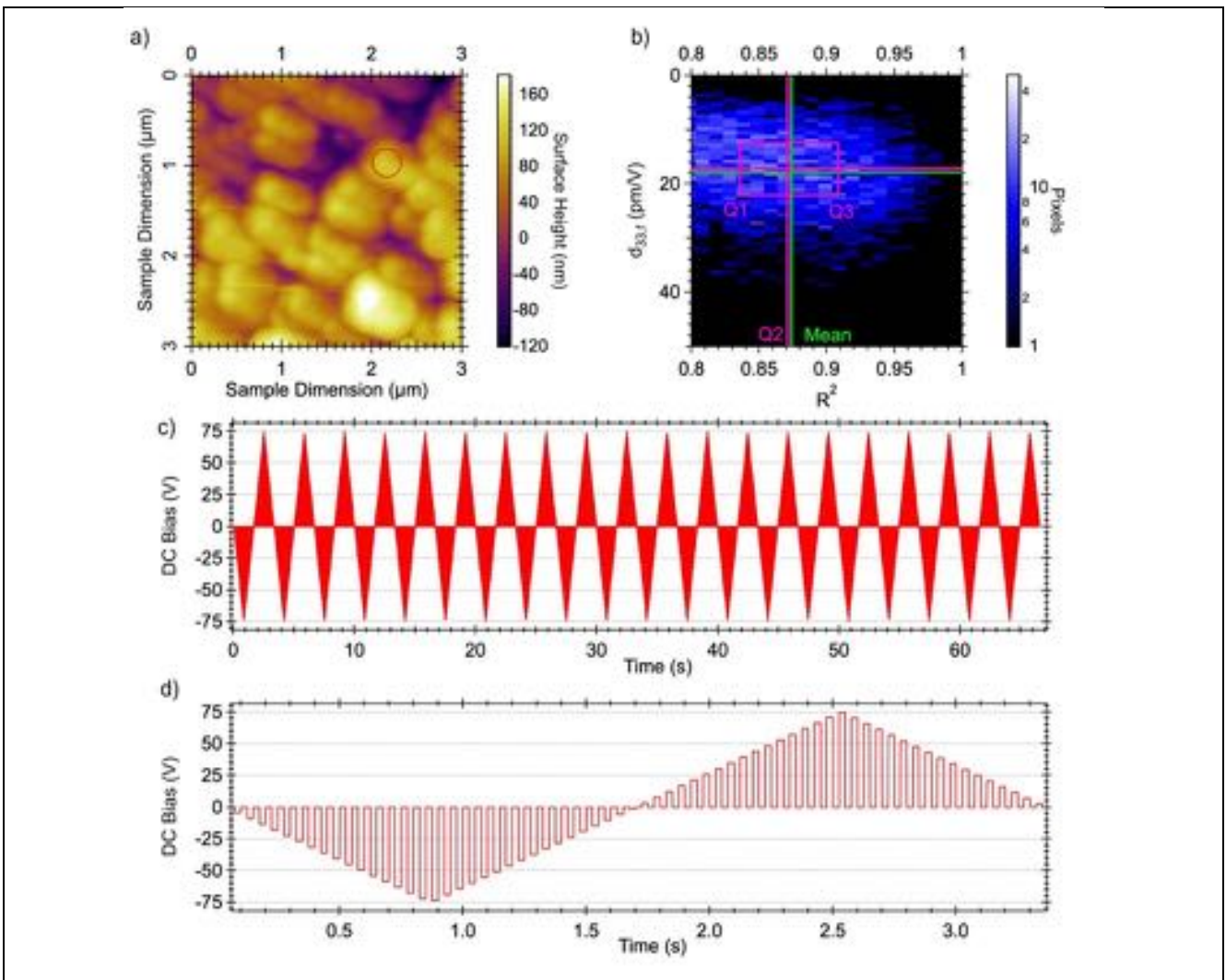

**Figure S.8.** Piezoresponse atomic force microscopy results from an as-deposited crystalline LaWN$_3$ thin film. (a) Surface height of a (3 μm)$^2$ scan, highlighting the location of the probe tip placement when the DC bias loading curve was applied. We note that due to the random polycrystalline microstructure of the sample, the switching is only observed for some grains and not others over this measurement range, which is to be expected for a polycrystalline film. (b) Pixel $d_{33,f}$ and $R^2$ distribution. (c) DC bias applied to test for ferroelectric character during PFM spectroscopy testing showing the (d) on and off states of a single DC loading cycle. A 5V AC piezoelectric drive voltage was applied in superposition to measure the piezoelectric response magnitude and phase during the DC loading and unloading cycles. Dual AC Resonance Tracking (DART)(7) controlled the AC piezoelectric drive frequency. Data shown in **Fig. 4f** and **Fig. 4g** were taken during the DC bias off (0V) states.

**Results: Piezoresponse Force Microscopy of PZT**

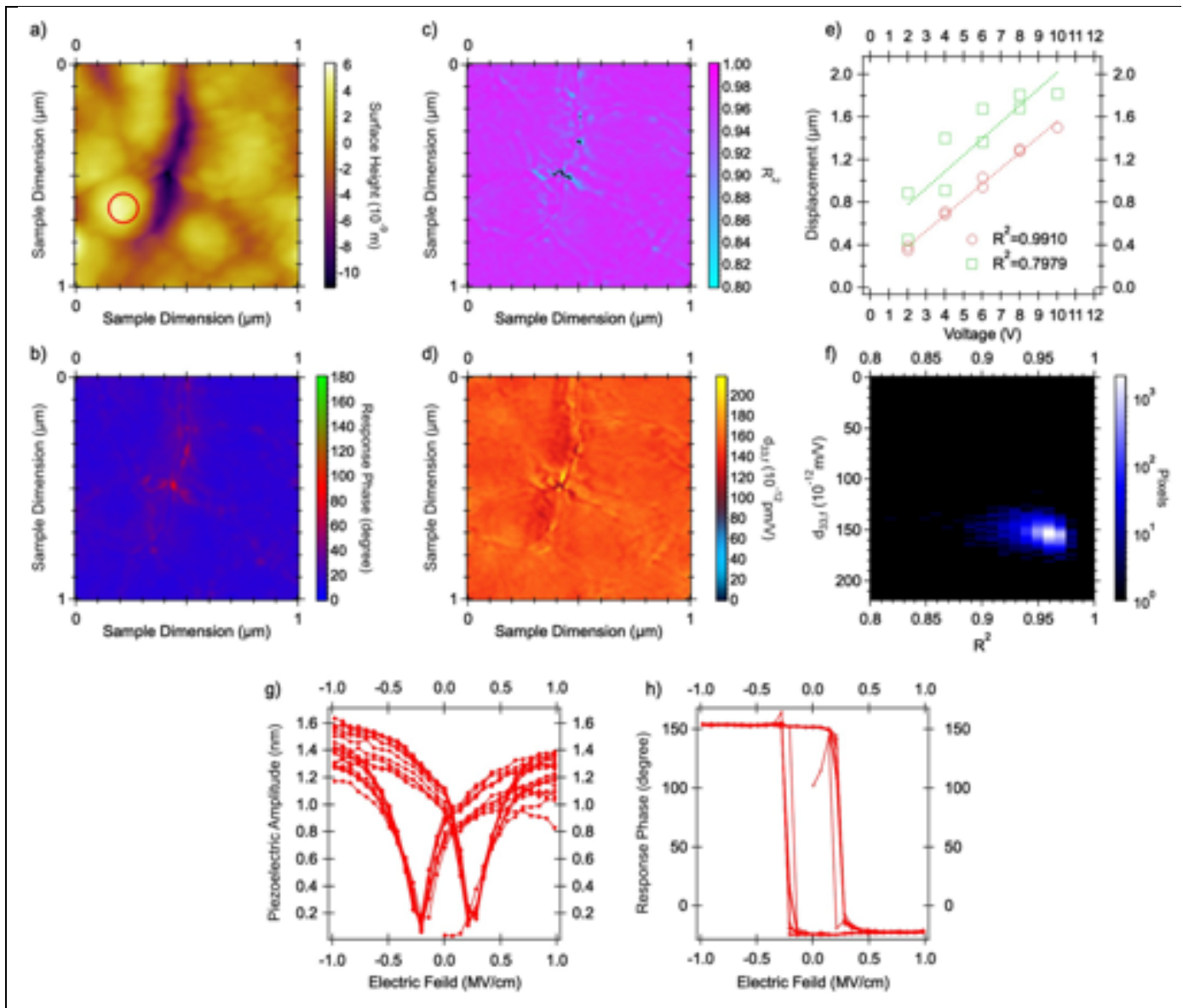

**Figure S.9.** Piezoresponse force microscopy results of a (111) textured PbZr$_{0.52}$Ti$_{0.48}$O$_3$ (PZT) thin film deposited on a (111) textured platinum substrate. (a) The surface is observed to be smooth with recessed grain boundaries. Piezoelectric displacement and (b) response phase as a function of drive voltage was collected in a dual frequency resonance tracking mode(7). Each frame pixel was fit with a line function to extract the (c) $R^2$ as a metric for response linearity and the slope as (d) effective piezoelectric strain coefficient ($d_{33,f}$) for each of the >16k pixels. The slope and linearity of each fit is shown for pixels where $R^2$>0.8, with (e) examples of best and worst linearity. (f) A 3D histogram of all $d_{33,f}$ and $R^2$ values shows a narrow distribution of fit values. The results indicate PZT has a clear piezoelectric response and reinforce the strength of this analysis method. (g) Piezoelectric response "butterfly loop" and (g) Hysteresis in response phase results when a DC bias loading curve is applied to the location highlighted in (a) and repeated through 10 cycles.

# Results: Piezoresponse Force Microscopy of AlScN

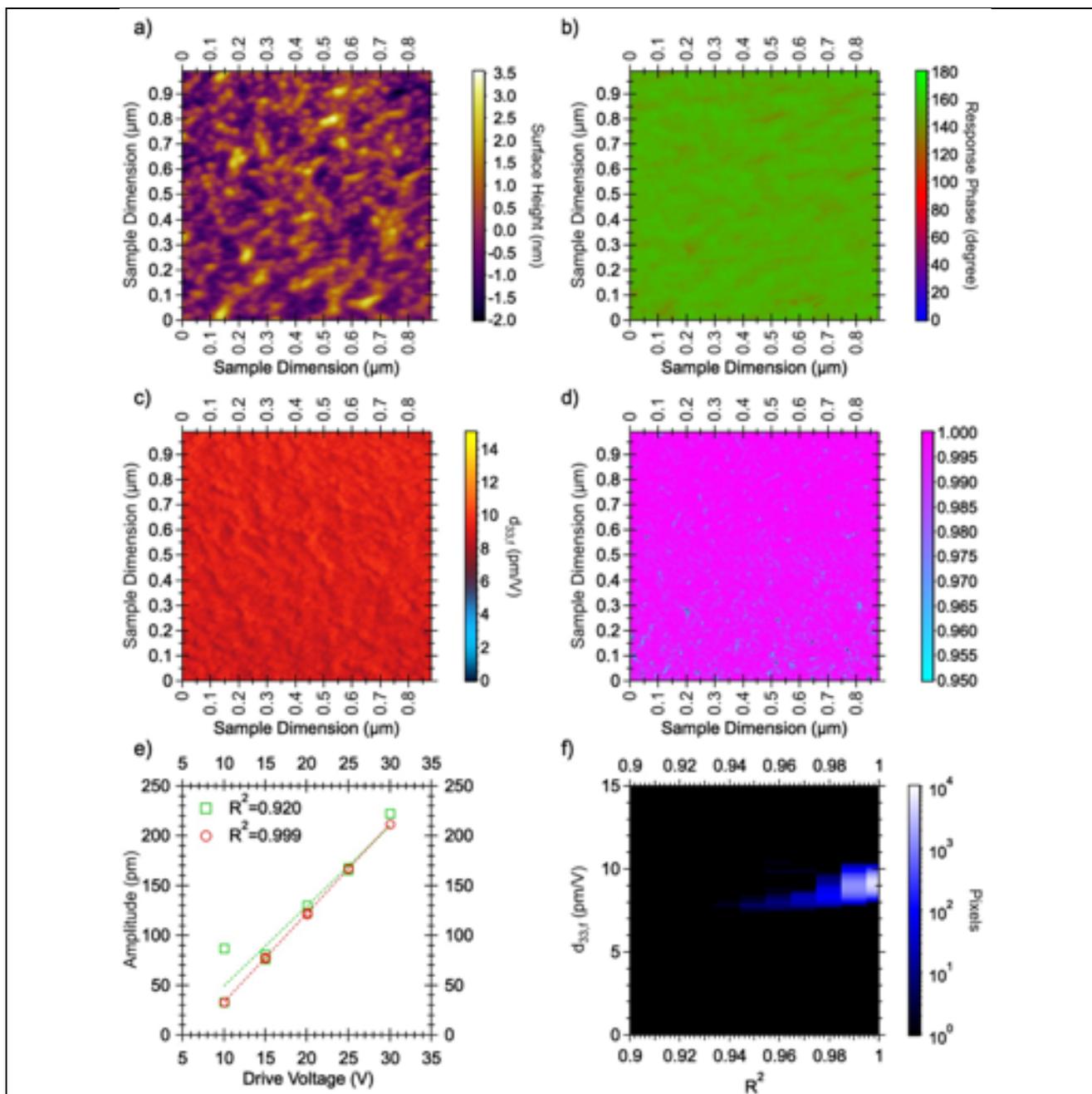

**Figure S.10.** Piezoresponse force microscopy results of a (002) textured $Al_{0.92}Sc_{0.08}N$ (AlScN) thin film deposited on a *p*-type (100) substrate. (a) The surface is observed to be smooth with fine grain structures. Piezoelectric displacement and (b) response phase as a function of drive voltage was collected in a dual frequency resonance tracking mode(7). Each frame pixel was fit with a line function to extract the (c) slope as the effective piezoelectric strain coefficient ($d_{33,f}$) and (d) $R^2$ as a metric for response linearity for each of the >16k pixels. The slope and linearity of each fit is shown for pixels where $R^2$>0.9, with (e) examples of best and worst linearity. (f) A 3D histogram of all $d_{33,f}$ and $R^2$ values shows a narrow distribution of fit values. The results indicate AlScN has a clear piezoelectric response and reinforce the strength of this analysis method.

**Discussion: Minimizing W and WN impurities**

In producing capped amorphous films and crystallizing post-deposition through thermal treatment, a second phase of metallic tungsten is observed with the perovskite structure. Similar behavior was observed in the oxynitride version, $LaWO_{0.5}N_{2.5}$ by the same authors.(15) This likely results from sub-stoichiometric nitrogen concentrations in the amorphous film, as seen in **Fig. 2d**, whereupon crystallization of the perovskite, excess tungsten and lanthanum build up at the grain boundaries and the higher mobility of smaller tungsten atoms in the grain boundaries leads to formation of body-centered cubic tungsten crystallites. The phase fractions resulting from the refinement shown in **Fig. 3b** are shown with various units in **Table. S.4**. It has been observed here that depositing with a substrate temperature of about 700°C results in films that are crystalline and oxidation-resistant. In the case of films deposited with heating, excess tungsten becomes rocksalt WN. For both deposition routes, excessively lanthanum rich materials result in amorphous La-W-N, which suffers from catastrophic oxidation. These pathways for producing crystalline $LaWN_3$ demonstrate the need for additional characterization and process optimization while also opening the door to further applications.

**Table S.4.** Refined phases of metallic tungsten (W) and lanthanum tungsten nitride ($LaWN_3$) shown in **Fig. 3b** and **Fig. S.2**. Conversions used standard atomic masses, in addition to the density and weight fractions resulting from the structural refinement shown in **Fig. 3b**.

| Phase | Density (g/cm$^3$) | Weight % | Molar % | Atomic % | Volume % |
|---|---|---|---|---|---|
| W (BCC) | 19.26 | 10 | 18 | 4 | 5 |
| $LaWN_3$ (R3c) | 9.47 | 90 | 82 | 96 | 95 |

**Supporting Information References**